\documentclass[aps,superscriptaddress,twocolumn,amsfonts,prl]{revtex4-2}
\usepackage{graphicx}
\usepackage{array}
\usepackage{amsmath,amssymb}
\usepackage{xcolor}
\usepackage{physics}
\usepackage{soul}
\usepackage{siunitx}
\usepackage{makecell}
\usepackage{hyperref}
\usepackage{dcolumn}
\usepackage{bm}
\usepackage{wrapfig}
\usepackage[normalem]{ulem}
\newcommand{\nSW}{n_{\text{SW}}}
\newcommand{\xx}{\mathbf{x}}

\newcommand{\BC}[1]{\textcolor{teal}{#1}}
\preprint{APS/123-QED}

\begin{document}

\title{Kin-ematic Exclusion in Active Matter: Modelling Mutual Inhibition in \textit{Pseudomonas aeruginosa} Sibling Colonies}

\author{Dario Buonomo}
\author{Francesco Imperi}
\affiliation{Science Department, Roma Tre University, Rome, Italy}
\author{Fabio Bruni}
\affiliation{Science Department, Roma Tre University, Rome, Italy}
\author{Marco Polin}
\email[]{email: mpolin@imedea.uib-csic.es}
\affiliation{Mediterranean Institute for Advanced Studies, IMEDEA, UIB-CSIC, Esporles, 07190, Spain}
\author{Barbara Capone}
\email[]{email: barbara.capone@uniroma3.it}
\affiliation{Science Department, Roma Tre University, Rome, Italy}

\date{\today}

\begin{abstract}
The striking variety of macroscopic morphologies displayed by bacterial colonies depends on microscopic environmental and behavioural details in a manner that is currently not well understood. A surprising example is sibling inhibition, whereby isogenic bacterial colonies spreading in soft agar hydrogels tend to avoid each other and form sharp demarcation lines when growing nearby. Here we investigate this effect with the common pathogen \textit{Pseudomonas aeruginosa}, by combining quantitative density measurements with a minimal biophysical model. Our results show that the phenomenon does not depend on gel compression, lethal inhibition or quorum sensing-dependent cell communication. Instead, colony separation is driven by localised nutrient depletion through a dynamic feedback between growth and motility. The model, which is calibrated using experimental data, captures key observations including the dependence of inhibition strength on the initial nutrient concentration. This work establishes nutrient availability and non-lethal motility inhibition as central factors underlying sibling inhibition, providing a generalisable framework for microbial spatial dynamics with implications for understanding bacterial interactions in tissues, soils and engineered microbiomes. 
\end{abstract}

\maketitle

Contributing approximately $15\%$ of Earth's biomass, bacteria underpin the productivity and biogeochemistry of all ecosystems, with significant impacts on the dynamics of trophic levels and on balances of global biogeochemical cycles \BC{\cite{bar-on2018}}.  These macroscopic effects depend on events at the micro-scale, from  biochemical transformations to interactions among cells  and with the surrounding microenvironment \cite{farooq07}. The microenvironment, in particular, is often chemically and physically heterogeneous at the bacterial scale. This is true in the ocean, where nutrients are usually concentrated in ephemeral micrometric hotspots \cite{stocker12} but even more so in solid media like soil sediments, foams or animal tissues, which often present a complex microscopic porous structure \cite{battacharjee21}. Understanding the peculiarities of bacterial motility, replication and colonisation within porous materials has therefore important biological, ecological and medical consequences \cite{berg_wolfe89,datta13,ribet_cossart15,toley_forbes11}. Here we investigate sibling inhibition, a puzzling phenomenon whereby bacteria prevent themselves from completely colonising an otherwise accessible porous material.

Bacterial motility within a porous structure is deeply affected by the size of pores \cite{croze11}. The classical run-and-tumble-like dynamics typical of bulk liquids changes into a hopping-and-trapping motion \cite{licata16,bassu24} with chemotaxis resulting from biases in hopping orientation rather than tumbling frequency modulation \cite{battacharjee21}.
Motility and replication allow colonies to expand within the porous structure, leading to a variety of interactions between neighbouring colonies, from competition for resources and invasion, to coexistence and cooperation \cite{gibbs08,stefanic15,lyons16,ternado-yuste2025,porter2025}. 
In some instances, neighbouring colonies do not merge despite their cells being motile, resulting in the formation of distinct macroscopic demarcation lines (DLs). 
This phenomenon is typically observed between genetically diverse isolates \cite{lyons16, patra17, kastrat24}, and in some cases it is employed for low-tech tests of genetic diversity within a clinical setting \cite{munson02}.
However, it has also been reported for sibling (clonal) colonies \cite{espeso16,beer09,kastrat24,sekowska09,rajorshi19}. This surprising phenomenon, termed sibling inhibition, does not depend on changes in bacterial behaviour in response to inter-species interactions, and could therefore be argued to be the simplest type of interaction between any two colonies coming into contact~\cite{Curatolo_2020}.
Three main mechanisms have been proposed to explain sibling inhibition. 
The first relies on matrix compression by the expanding colonies, leading to smaller pore sizes. If pores in the region between two fronts are compressed enough, they could become too small for cells to swim through, creating an inaccessible region~\cite{espeso16}.
The second depends on the production of an inhibitory molecule, either lethal \cite{beer09,beer10,kastrat24} or non-lethal \cite{cruz-lopez21}, which causes cells to die or stop moving above a threshold concentration. This might depend on quorum-sensing pathways, and can lead a sub-population of cells to lyse and release nutrients.   
The third mechanism relies on cell motility increasing for decreasing nutrient concentration. Coupled with nutrient diffusion and its consumption by cells, this would lead  neighbouring colonies to merge when grown on low nutrient concentration, and form clear DLs otherwise \cite{sekowska09}. 
Although all of these effects could in principle play a role in the formation of DLs, assessing their importance is complicated by the variability in species and physico-chemical growth conditions used so far in the literature, as well as the types of observables reported. As a result, a clear consensus has yet to emerge.

Here we tackle this problem by studying sibling inhibition in the Gram-negative bacterium \textit{Pseudomonas aeruginosa}, a dangerous human pathogen frequently associated with nosocomial infections \cite{cervoni23,sposato24,cervoni25,lichtenberg22}. 
Through a systematic set of experiments, we provide a fine-scale description of the macroscopic expansion dynamics of colonies --including DL formation-- and microscopic cell motility and cell replication properties. Combining experiments with mathematical modelling, we test the main hypotheses for DL formation and show that, in the case of \textit{Pseudomonas aeruginosa}, sibling inhibition results mainly from nutrient depletion leading to a progressive decrease in both replication rate and cell diffusivity. 

\begin{figure}[t]
    \includegraphics[width=0.95\linewidth]{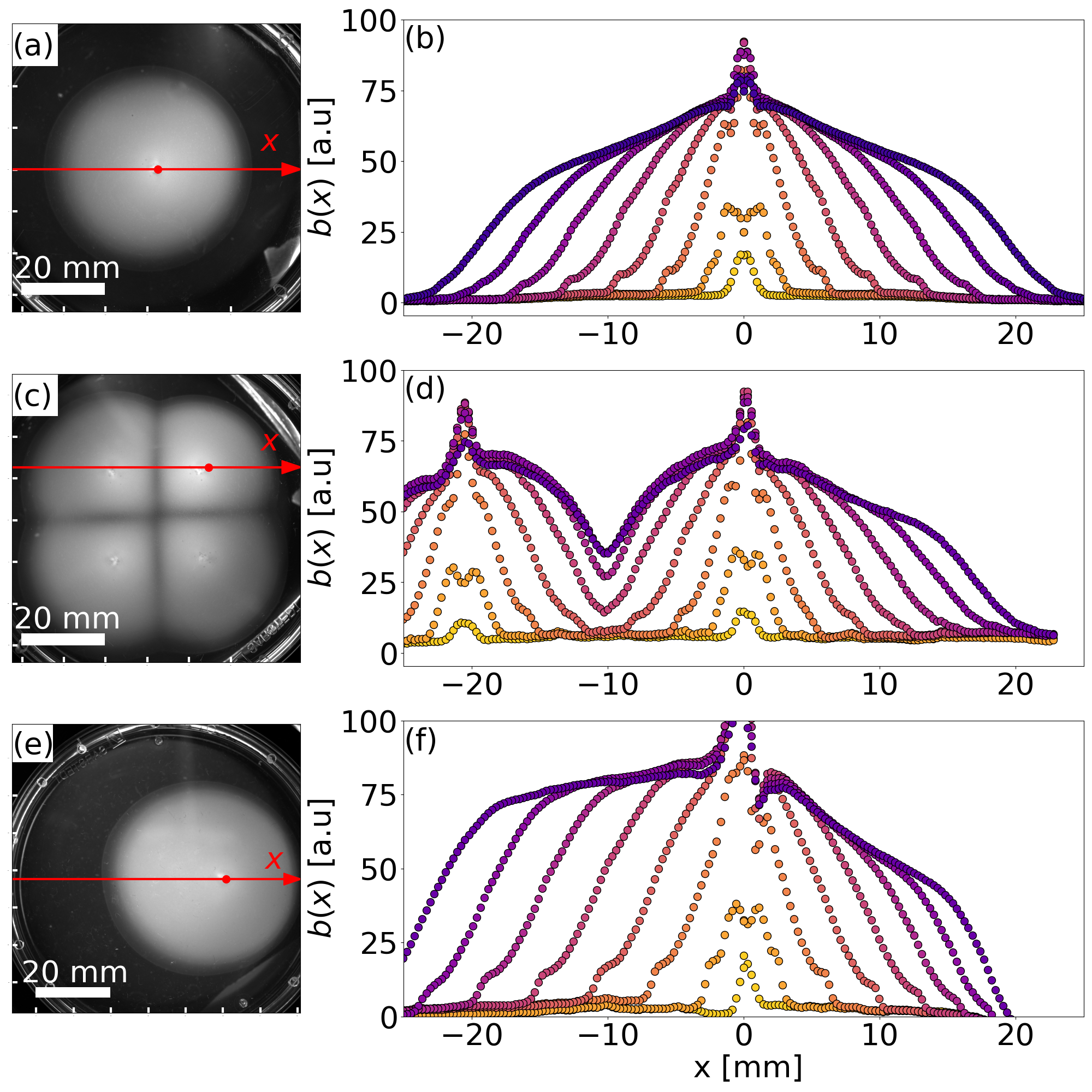}
    \caption{ {\bf Spreading of PAO1 colonies in SW agar} ($n_0=1$). 
    a) Individual colony $24\,$h post inoculation. Inoculation point at the centre of the plate.
    b) Evolution of the bacterial density profile for the colony in a). Curves are azimuthally averaged around the inoculation point ($x=0$). 
    c) Four colonies $24\,$h post inoculation. The demarcation lines are clearly visible. 
    d) Evolution of the bacterial density profile for the colonies in c). Profiles are calculated along the four lines marked in panel c), with $x=0$ a reference inoculation point. The profiles are then averaged. 
    e) Individual colony $24\,$h post inoculation. The inoculation point corresponds to one of the positions used in c). 
    f) Evolution of the bacterial density profile for the colony in e).}     
\label{fig_01}
\end{figure}

\section{Methods}

\subsection{Preparation of bacterial stocks}
Cultures of {\it Pseudomonas aeruginosa} (wild type strain PAO1 and isogenic mutant strains) were prepared from frozen stocks by streaking on standard LB $1.5\%$~w/v agar plates. Plates were then incubated at $37^{\circ}$C overnight and stored at $5^{\circ}$ for up to 3 weeks, to be used as stocks for subsequent experiments. 

\subsection{Swimming agar assays}
Swimming agar experiments were performed by stab inoculation of soft agar plates ($0.3\%$~w/v), based on a low nutrient LB-like swimming medium (SW) containing $0.5\%$~w/v NaCl, $0.1\%$~w/v Tryptone and $0.05\%$~w/v yeast extract. 
In experiments, we varied the background nutrient concentrations ($n_0$) with respect to this reference medium ($n_{\text{SW}}$), keeping the salt concentration constant.
These agar concentrations are known to be able to mimic the physical property of human mucosae, like the  internal mucosa of lung tissues \cite{vinod23}, where the most dangerous \textit{P. aeruginosa} infections usually take place. 

After inoculation, a custom-built imaging setup was used to record colony expansion at $2\,$min intervals for 24 hours.

\subsection{Concentration of viable bacteria}
The local concentration of viable bacteria in swimming experiments was quantified through the colony forming unit (CFU) assay via the standard plate count method. Local samples from the primary culture plate 
were serially diluted tenfold in sterile saline, and four 5 $\mu$L aliquots of each dilution were spotted onto LB $1.5\%$ agar plates. Individual colonies were counted after overnight incubation at $37^{\circ}$C. Each sample was repeated in four replicates.

\begin{figure}[t]
    \centering
    \includegraphics[width=\linewidth]{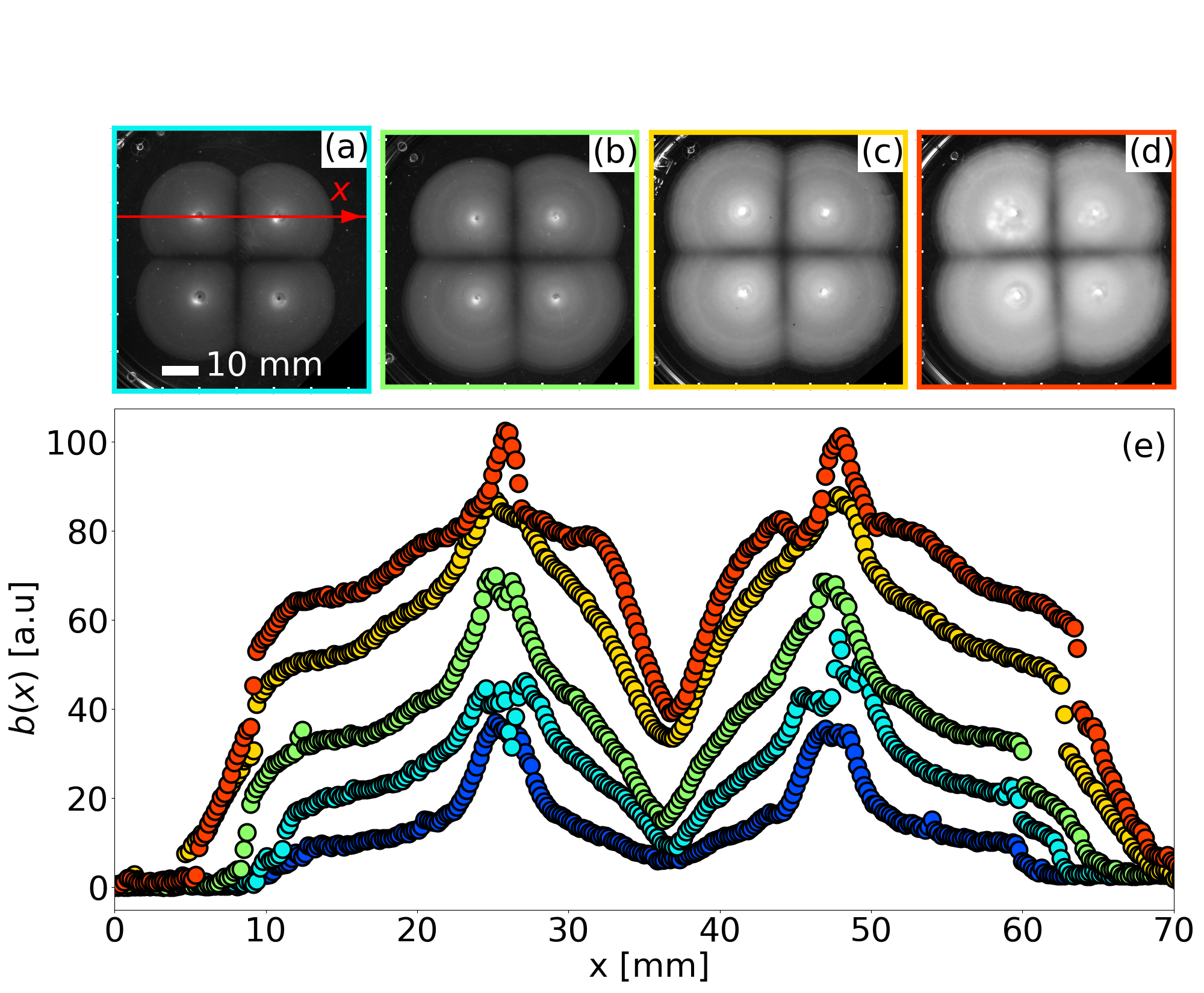}
    \caption{{\bf Demarcation lines vs. background nutrient concentration.} a-d) Plates with four colonies of PAO1, $16\,$h post inoculation. They correspond respectively to background nutrient concentrations $n_0=0.5, 1, 2, 3$. The scale bar is $10\,$mm.
    e) Bacterial density profiles at $24\,$h for $\nSW$ equal to $0.1$ (blue circles), 0.5 (cyan circles),  1 (green circles),  2 (yellow circles), and  3 (orange circles). The image of $n_0 = 0.1$ at $24\,$h is not included because the colonies do not expand sufficiently to either merge or give rise to a clear DL.  All profiles were obtained similarly to those of Fig.~\ref{fig_01}d.}
    \label{fig_02}
\end{figure}

\subsection{DDM estimates of motility parameters for planktonic cells}
Differential dynamic microscopy (DDM), first proposed in \cite{cerbino08}, is a widely used technique to study the dynamical properties of colloidal suspensions and active agents like motile bacteria \cite{SCHWARZLINEK20162, wilson11}. For bacterial suspensions, under mild assumptions, DDM can be used to extract the mean swimming speed $\bar{v}$, the standard deviation of the velocity distribution $\sigma$, the fraction of motile cells $\alpha$ and the (thermal) diffusion coefficient of non-motile ones $D$ \cite{Germain_DDM, Martinez_DDM}. For an exhaustive introduction, we refer the interested reader to \cite{lattuada25}. Briefly, starting from a low-magnification movie of a bacterial suspension, DDM uses the Fourier Transform of individual images,  $\tilde{I}(\mathbf{q},t)$ to calculate the intermediate scattering function (ISF) 
\begin{equation}
 \label{eq:ISF_defo}
     f(q,\tau)=\frac{   \langle  \tilde{I}^{*}(\mathbf{q}, t) \tilde{ I}(\mathbf{q}, t+\tau ) \rangle}{    \langle  | \tilde{ I}(\mathbf{q}, t) |^{2} \rangle},
 \end{equation}
where $*$ indicates complex conjugation and the averaging runs over $t$ and the direction of the wavevector $\mathbf{q}$. For bacterial suspensions, at short timescales $f(q,\tau)$ can be approximated as \cite{Martinez_DDM}:
\begin{equation}
f(q,\tau)= e^{-q^2 D \tau}(1- \alpha) + \alpha e^{-q^2 D \tau} \mathcal {P}(q, \tau). 
\label{eq:ISF_bacteria}
\end{equation}
Here $\mathcal {P}(q, \tau)$ describes the contribution of motile bacteria to the intermediate scattering function. It reads
\begin{equation}
\label{eq:ISF_P_for_motile}
\mathcal {P}(q, \tau) = \int_{0}^{\infty} P(v)\frac{\sin(qv\tau)}{ qv\tau} dv.
\end{equation}
As usual, the probability distribution function of swimming speeds, $P(v)$, is taken here to be a Shulz distribution with mean $\bar{v}$ and variance $\sigma^2$ (see Eq.~9 in \cite{wilson11}).  The four motility parameters are estimated by fitting the experimental intermediate scattering function to its theoretical form in Eq.~\ref{eq:ISF_bacteria}. 
For the experiments, we begin by picking a colony from the stock plate and incubate it in fresh liquid LB medium at $37^{\circ}$C for 5 hours under constant shaking at $120\,$rpm. Cells are then washed three times with dH2O by centrifuging for $5\,$min at $3\,$rcf and substituting the supernatant. 
Tubes with $100\,\mu$l liquid media at different nutrient concentrations are then inoculated with $100\,\mu$l of washed cell suspension and left to acclimate for $30\,$min. 
The liquid media are prepared in the desired conditions following dilution.
Observation chambers $80\,\mu$m-thick were loaded with the bacterial suspension and immediately imaged at $10\times$ (Olympus Plan N $10\times$ objective with NA 0.25 ) with a Olympus IX81 inverted microscope, equipped with a Thorlabs Zelux  CS165CU1(/M) CMOS camera.
We typically recorded bacterial motility for $70\,$sec at $70\,$fps.

\subsection{Growth experiments}
For each planktonic growth experiment, three samples were picked from different colonies on the same stock plate, and incubated in fresh LB medium at $37^{\circ}$C for 8 hours under constant shaking at $120\,$rpm. Samples were then loaded in triplicates in 96-well plates filled with test media, at a dilution factor of $10^3$.  We incubated the plates in a multimode micro-plate reader (Spark, Tecan, Switzerland), which monitored the growth in individual wells for 14 hours. Before each measurement, the plates were prepared by shaking them at $180\,$rpm for $10\,$s. Reference wells containing only  media provided a sterility control and were used to set the background signal. 
For each growth condition, curves from all samples and replicates were used to estimate the experimental growth dynamic and its variability. 

\begin{figure}[t]
    \centering
    \includegraphics[width=\linewidth]{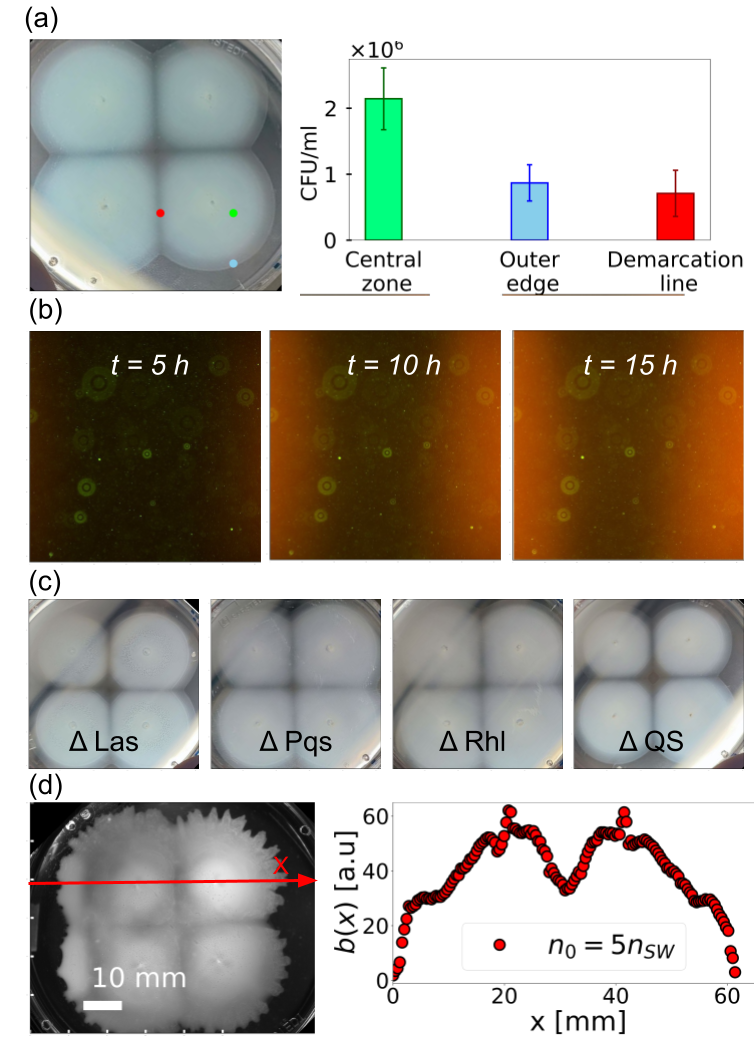} 
    \caption{{\bf Potential mechanisms generating DLs.} a) \ul{Lethal inhibition}. Sampling regions and corresponding CFU counts.
b) \ul{Matrix compression}. Green fluorescent colloids embedded in the gel matrix as a colony of red-fluorescent {\it P. aeruginosa} cells advances across the field of view. The three panels correspond respectively to $5, 10, 15\,$h post inoculation.
c) \ul{Quorum sensing}. Colony profiles $24\,$h post-inoculation for $\Delta$Las, $\Delta$Pqs, $\Delta$Rhl, $\Delta$QS. 
d) \ul{Nutrient levels}. Colony profile $24\,$h post-inoculation for $n_0=5$. Notice the frayed outer edge of the colony. The plotted density profile corresponds to the solid line marked on the left.
}    
    \label{fig_03} 
\end{figure}

\subsection{Diffusion simulations}
The partial differential equations (PDEs) modelling the spatio-temporal evolution of bacterial concentration, $b(\mathbf{x},t)$, and nutrient concentration, $n(\mathbf{x},t)$, were simulated on a 2D grid of size $1000\times 1000$ ($10\times 10\,$cm) with a custom Python code. We used an explicit finite difference method and a fixed time step of $6\,$s, for a total of 24 simulated hours. The time step was chosen to ensure the stability of the explicit scheme used for PDE integration.
All simulations were initialised with a uniform nutrient concentration, $n(\mathbf{x},t=0)=n_0$, and Gaussian profiles with standard deviation $500\,\mu$m for the stabs (see section S3 of the Supplementary Materials for more details).

\subsection{Fluorescence experiments}
Red fluorescent \textit{Pseudomonas aeruginosa} (recombinant PAO1 derivative ectopically expressing mCherry \cite{popat12}) was cultured as described above. Standard swimming agar plates were prepared with the addition of green fluorescent silica colloids ($\sim1\,\mu$m in diameter). Fluorescent bacteria were inoculated onto the agar plates in a grid pattern, and the midpoints between inocula were imaged overnight at $10\times$ magnification using a Nikon Ti2 inverted microscope equipped with an incubation chamber maintained at $37^\circ$C. Both red and green fluorescent signals were recorded throughout the experiment to confirm the absence of any net displacement of GFP-tagged colloids during bacterial expansion in the same field of view.

\section{Results}

\begin{figure}[t]
    \centering 
    \includegraphics[width=\linewidth]{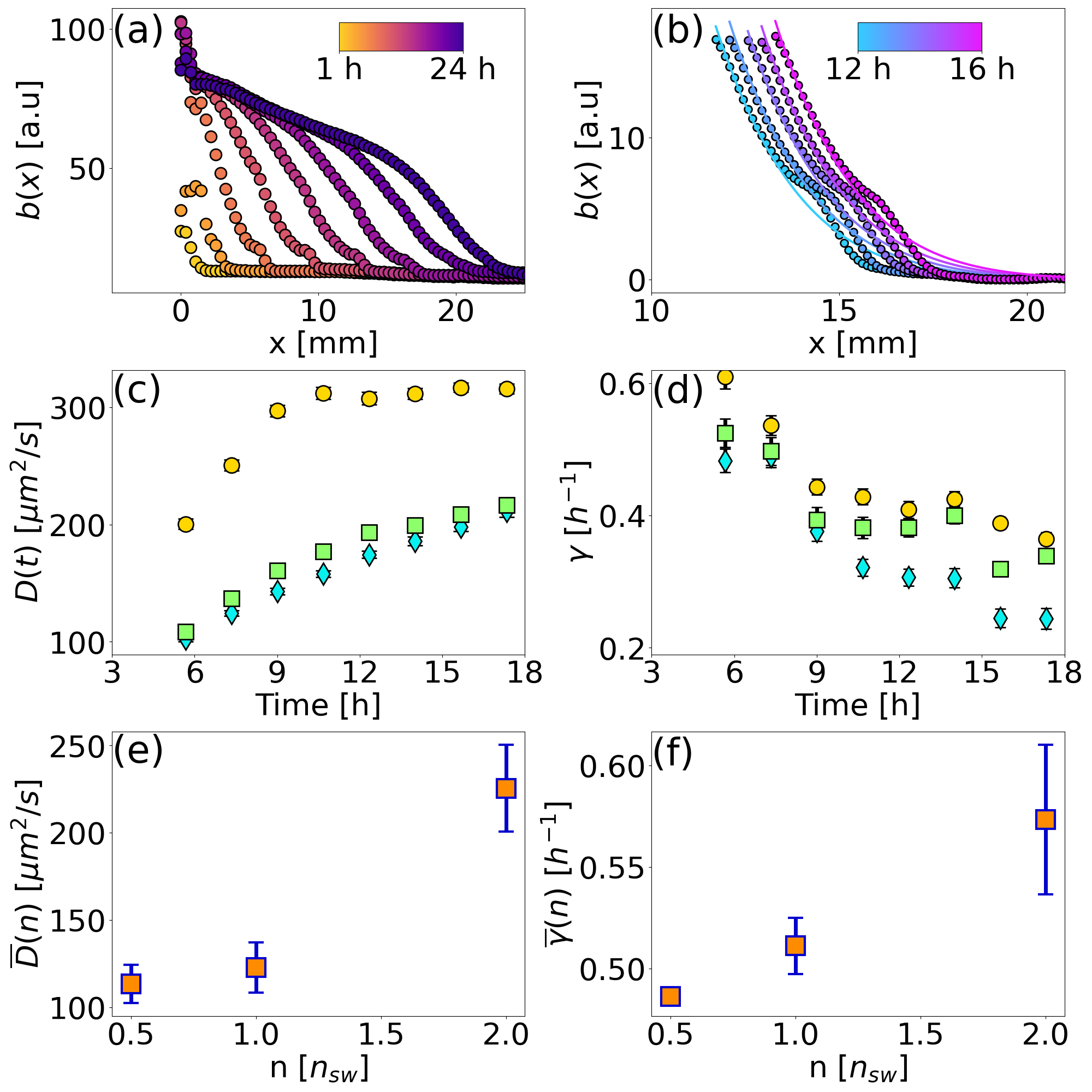}
    \caption{ {\bf Dynamics of single colony expansion.} a) Evolution of the azimuthally averaged radial density profile $b(\rho,t)$ for a single colony of PAO1 spreading on SW agar. The profiles are displayed here every $3\,$h over a  period of $24\,$h post inoculation. b) Tails of $b(\rho,t)$ used for the fits to Eq.~\ref{eq:multifit}. 
c,d) Time evolution of the diffusivity (c) and replication rate (d) for 3 different nutrient levels ($n_0=0.5,1,2$; cyan, green, yellow respectively).
e,f) Dependence of $D(n)$ (e) and $\gamma(n)$ (f) on $n_0$. The values plotted correspond to the first time point in the temporal evolution plotted above. 
    }
    \label{fig_04}
\end{figure}

\subsection{{\it P. aeruginosa} DLs do not depend on lethal inhibition, gel compression or quorum sensing, but change at high nutrient concentration}
Figure~\ref{fig_01} shows the typical spreading dynamics of PAO1 in SW agar plates ($n_0=1\nSW$), starting from either a single colony (Fig.~\ref{fig_01}a,b) or 4 sibling colonies (Fig.~\ref{fig_01}c,d). 
As expected, an individual colony spreads isotropically around the inoculation point, although the boundary of the Petri dish can lead to anisotropies if the initial inoculum is off-centre (see Fig.~\ref{fig_01}e,f).
When multiple clonal colonies are present, however, clear demarcation lines develop with time (Fig.~\ref{fig_01}c,d). 
This phenomenon is present for a wide range of nutrient concentrations ($0.1<n_0<3$, Fig.~\ref{fig_02}), and persists for more than $24\,$h. 

In order to rationalise the development of DLs, we proceeded to test the mechanisms described previously in the literature.
Beginning with lethal inhibition, reported for \textit{Paenibacillus dendritiformis } swarms \cite{beer09,beer10}, we incubated $3$ plates with four sibling colonies for $24\,$h, and then estimated the CFU density at three positions for each plate (see Fig.~\ref{fig_03}a), one well within the colony (central zone), one at the outer edge of the colony (free edge) 
and the last at one of the DLs. 
Figure~\ref{fig_03}b shows that, as expected, the central region has a significantly higher concentration of viable cells than the colony edges. However, there did not appear to be any noticeable difference between the two edge samples. Cells along the demarcation line were characterised by the same viability as along any other parts of the colony edge. This shows that, differently from the case of \textit{P. dendritiformis} swarms, viability loss is not responsible for the development of the DLs we observe. 
The emergence of DLs between {\it Pseudomonas putida} colonies growing in swimming agar has been proposed to originate from compression of the gel matrix by the expanding edges of the colonies  \cite{espeso16}.  Compression reduces the average pore size within the gel, hindering active bacterial diffusion and physically preventing cells from accessing the region between two colonies. 
In order to test for this in our case, we embedded tracer colloids in SW agar plates (10 replicates), and monitored the evolution of their displacement while sibling colonies were growing. The results are exemplified in Fig.~\ref{fig_03}c, showing the tracers' position at the midpoint between two expanding colonies. Here, as well as in all other experiments, we did not observe any compression of the gel as the colonies expanded and formed DLs. We therefore discard also gel compression as a cause for DL formation in our system. 
Finally, we tested whether any quorum sensing (QS) dependent factor could influence the formation of the DLs. \textit{P. aeruginosa} has three main quorum sensing systems, Las, Rhl and Pqs, each relying on distinct signalling molecules.
To evaluate the contribution of quorum sensing to colony avoidance, we tested mutants defective in the production of quorum sensing signalling molecules either separately (referred to as $\Delta$Las, $\Delta$Pqs, or $\Delta$Rhl), or together ($\Delta$QS) \cite{rampioni10, letizia22, mellini23, bondi17}. Figure~\ref{fig_03} shows that all mutants exhibited an identical inhibition behaviour even in the $\Delta$Las and $\Delta$QS mutants, which exhibit a minor reduction in spreading speed.
This suggests that DL formation in \textit{P. aeruginosa} does not rely on quorum sensing.

Having shown that the DLs we observe do not result from lethal inhibition, gel compression or quorum sensing, we now turn to the effect of nutrient availability. Figure~\ref{fig_03}j shows that, although DLs are always present for background concentrations $n_0\leq3$, higher concentrations (here $n_0=5$) lead to much less defined avoidance regions, which are only present within the inner area between the four colonies. Previous studies with another bacterial species \cite{sekowska09} showed that sibling inhibition progressively disappeared when lowering the background nutrient concentration,  due to an increase in cell swimming speed. Although here we observe the opposite phenomenology, DLs disappearing when nutrient concentration increases, these results suggest that the cells' response to nutrient availability could be responsible for sibling avoidance in our system.

\subsection{The bacteria-nutrients model}

In order to study the dynamics of the experimental system, we turn to a simple minimal model combining the dynamics of the bacterial concentration $b(\xx,t)$ and nutrient concentration $n(\xx,t)$. Bacteria spread with an active diffusivity $D(n(\xx,t))$ and grow at a rate $\gamma(n(\xx,t))$, both functions of the local background nutrient concentration. The nutrients diffuse with a constant diffusivity $D_n$ and are consumed by bacteria at a rate $\alpha(n(\xx,t))$. Altogether this reads: 
\begin{equation}
\label{eq:full_system}
\begin{cases} 
\partial_t b(\xx,t) =  \nabla^2\left[D(n(\xx,t))b(\xx,t)\right]+\gamma(n(\xx,t))b(\xx,t)\\ 
\partial_t n(\xx,t) =  D_n\nabla^2n(\xx,t)  - \alpha(n(\xx,t))b(\xx,t),
\end{cases}
\end{equation}
where the space (and time) dependent diffusivity has been incorporated here according to the It\^o convention, as commonly done in active matter \cite{lau07,cremer19,battacharjee21} (see section S4 of the Supplementary Materials for a discussion on the choice of the convention).

\begin{figure}[t]
    \centering
    \includegraphics[width=\linewidth]{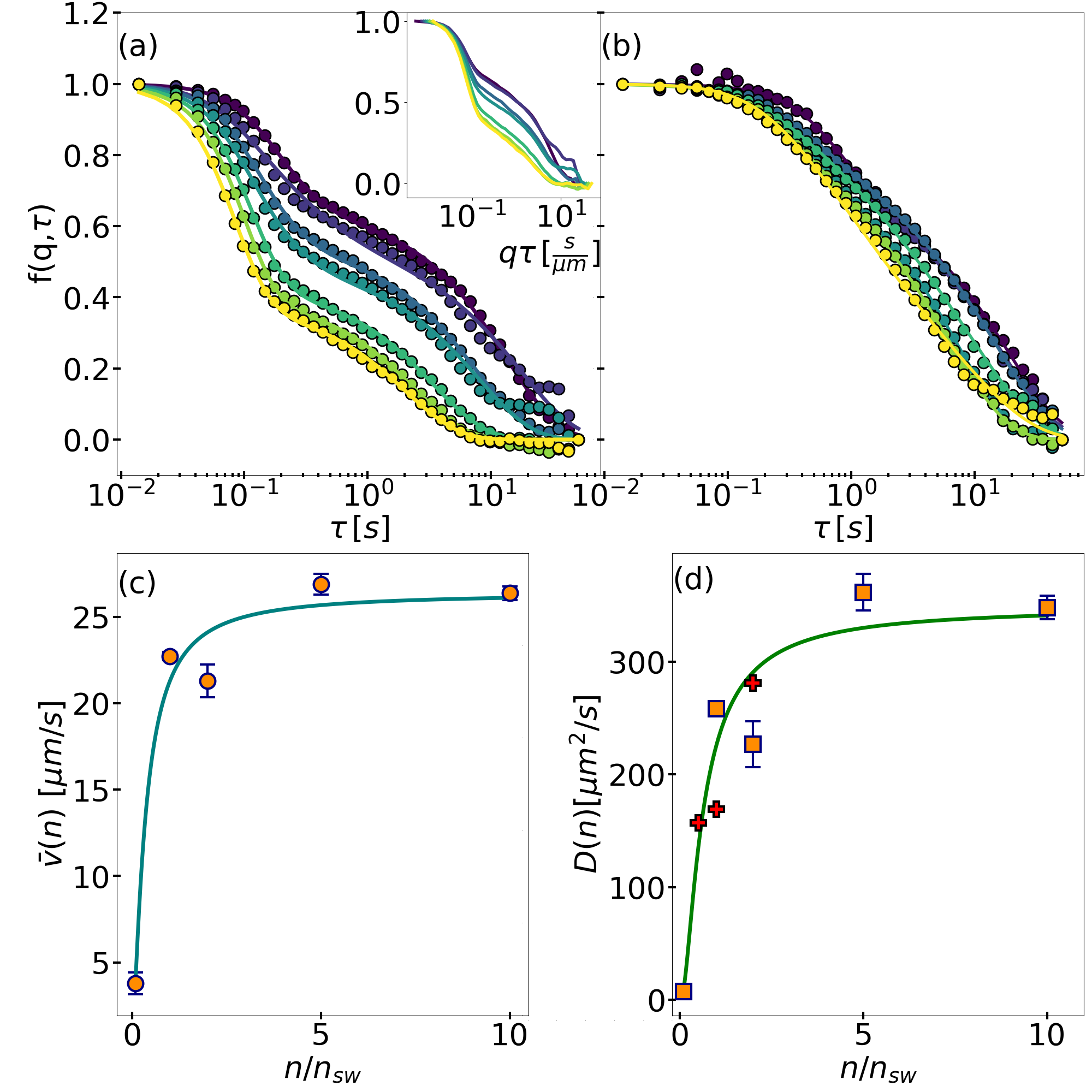}
    \caption{{\bf DDM of bacterial cultures.} 
a) Intermediate Scattering Function (ISF) for $n_0=10$ sample. Inset: ISF profiles plotted vs. $q\tau$. Curve collapse highlights the range of ballistic behaviour. 
b) ISF for $n_0=0.1$. For both panels a) and b) $q \in $[0.39 , 0.95] $\mu m^{-1}$.
c) DDM-derived average swimming speed of bacteria vs. background nutrient concentration (orange circles). The solid line is a fit to Eq.S8 of the Supplementary Materials. 
d) Effective diffusion coefficient corresponding to the velocities in panel c) (experimental data: red squares; heuristic fit: solid line; see text for details). Red crosses correspond to $D(n)$ estimates from Fig.~\ref{fig_04}.    
    }
    \label{fig_05}
\end{figure}

The dependence of bacterial diffusivity, replication rate and nutrient consumption rate on the background nutrient concentration are unknown {\it a priori} and need to be determined experimentally. 
To this end, we focus on the regions and times of the expansion dynamics for which gradients in nutrient concentration can be taken to be of the same order as gradients in bacterial concentration, or smaller. 
In general we expect this to be a valid assumption whenever bacterial concentration is smaller than a threshold value $b_{\text{th}}$ (here 30 [a.u]), like the initial stages of expansion or the outer edge of the expanding colony.
Within this approximation, $\nabla^2(Db)\simeq D\nabla^2b$ and the bacterial density can be expressed locally as a function of radius $r$ as
\begin{equation}
\label{eq:multifit}
b(r,t) \propto \exp\left({\gamma(n(r,t))t}\right)\frac{\exp \left(-\frac{r^2}{4D(n(r,t))t}\right)}{4\pi D(n(r,t))t}, 
\end{equation}
with an unknown proportionality factor that depends on time. The diffusivity and replication rate in this region can be estimated from the experimental expansion profiles (Fig.~\ref{fig_04}a). The outer edge of the full set of profiles captured within a $1\,$h interval centred at a given time $t$ is fitted to Eq.~\ref{eq:multifit} assuming a constant (but unknown) value of the nutrient concentration (Fig.~\ref{fig_04}b). The fit provides the evolution of $\gamma$ and $D$ at  the colony edge (Fig.~\ref{fig_04}c,d) as a function of time. Although the full curves cannot be used directly to read the dependence of replication and diffusion on $n$, for the initial time point we can still approximate the nutrient concentration by the background concentration used in the plates ($n\simeq n_0$). The results, shown in Fig.~\ref{fig_04}e,f show a positive correlation of both diffusivity and replication rate with nutrient concentration.

\subsection{Diffusivity vs. nutrient concentration}

As an independent test of how nutrient concentration affects bacterial motility, we next consider planktonic cultures.
 Following the protocol outlined earlier, we perform a series of DDM experiments at a range of background nutrient concentrations $0.1\leq n_0 \leq 10$. 
Figure~\ref{fig_05}a,b show representative plots of the intermediate scattering functions for background nutrient concentrations of 10 and 0.1 respectively, together with their fits to Eq.~\ref{eq:ISF_bacteria}. We see that the ISF model  describes well the experimental curves in both cases. Differently from cases reported previously \cite{sekowska09}, here the average swimming speed increases with nutrient concentration (Fig.~\ref{fig_05}c). The dependence can be captured heuristically by the function 
\begin{equation}
\bar{v}(n) = \frac{n^m}{\nu_{v}^m + n^m}v_{\text{max}},
\label{eq:DDMspeed}
\end{equation}
where the characteristic concentration $\nu_{v}=0.35\pm 0.14$, the exponent $m=1.4\pm 0.4$, and the maximum speed $v_{\text{max}}=26\pm 2\,\mu$m/s. The diffusivity of planktonic cells can be written as $D(n)\propto \bar{v}^2(n)\tau$, with a characteristic time scale $\tau$ typically given by the mean run time, here $\sim1\,$s , and  a proportionality factor of order 1 that depends on dimensionality \cite{martens12}.  

Although diffusion in porous media requires a subtler modelling strategy to link the microscopic cell dynamics with macroscopic diffusivity \cite{licata16,battacharjee21,bassu24}, for the loose swimming agar used here we expect the planktonic expression to still be valid. Figure~\ref{fig_05}d compares the diffusivity estimates from the colony expansion experiments (Fig.~\ref{fig_04}e) with the curve $D(n)=\bar{v}^2(n)\tau/2$ expected for 2D expansion ($\tau=1\,$s).  We see that the curve derived from DDM measurements agrees well with the colony expansion data, without fitting parameters. It will therefore be used to inform the colony expansion model.

\subsection{Replication and consumption rates vs. nutrient concentration}
\begin{figure}[t]
\centering  
\includegraphics[width=\linewidth]{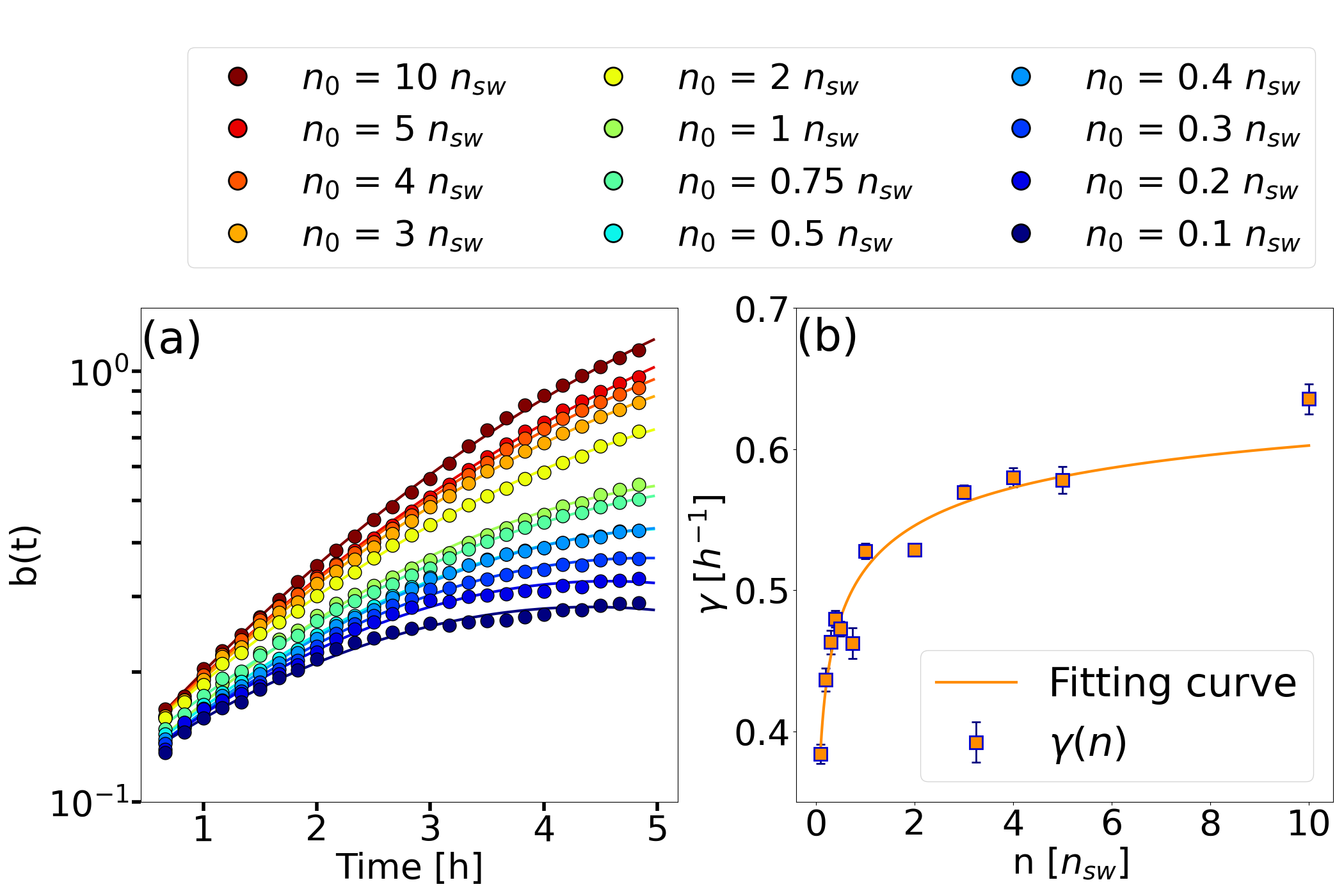}
\caption{{\bf Growth dynamics of PAO1 for different nutrient concentrations.}
a) Initial bacterial concentration growth vs. background nutrient concentration $n_0$ ($n_0=0.1, 0.2, 0.3, 0.4, 0.5, 0.75, 1, 2, 3, 4, 5, 10$ respectively, as the curves are ordered from the smallest to the largest). Solid lines are fits to Eq.~S8 of the Supplementary Materials .
b) Replication rate $\gamma(n)$ as a function of the initial nutrient concentration. Orange squares: experimental values. Solid line: fits to Eq.~\ref{eq:growth_rate}.
}
\label{fig_06}
\end{figure}

In order to estimate replication and consumption rates, we perform a set of growth experiments in liquid media. Following the protocol described previously, we monitor for $14\,$hours the growth of bacterial cultures in individual wells of a 96-well plate, as a function of initial nutrient concentration. Considering the system to be well-mixed, the concentrations of bacteria and nutrients will evolve according to 
\begin{equation}
\begin{cases} 
\partial_t b(t) =  \gamma(n(t))b(t)\\ 
\partial_t n(t) =   - \alpha(n(t))b(t) 
\end{cases}.
\label{eq:homogeneus_system}
\end{equation}
As discussed in detail in the Supplementary Material (Sec.~S2), under mild assumptions the initial part of the growth curve can be fitted to $\ln b(t) \simeq \ln b_0 + t\gamma(n_0) -\frac{t^2}{2}\alpha(n_0)b_0\frac{\partial \gamma}{\partial n}\Big|_{n_0}$ (see Fig.~\ref{fig_06}b), providing direct estimates of the growth rate $\gamma(n_0)$ as a function of the initial nutrient concentration $n_0$ in the well. 
The results are shown in Fig.~\ref{fig_06}c together with the best fit to a Moser growth model 
\cite{monod49, moser1957, Regis2026, jumpei25}
\begin{equation}
\gamma(n) = \frac{n^b}{\nu_{\gamma}^b+n^b}\gamma_{\text{max}}, 
\label{eq:growth_rate}
\end{equation} 
for which $\gamma_{\text{max}}=(0.6\pm 0.1)\,\text{h}^{-1}$, the characteristic concentration $\nu_{\gamma}=(0.12\pm 0.06)\,\nSW$ and $b =0.5 \pm 0.2$. 
The Moser growth model, an extension of Monod kinetics, provides a robust and widely employed  framework of bacterial growth dynamics under nutrient-limited conditions \cite{cremer19,battacharjee21}. Crucially, this model is predicted to emerge in multi-component media, such as the one employed here, where the sequential exhaustion of different chemical species imposes successive constraints on the specific growth rate \cite{yamagishi25}.

\begin{figure}
    \centering
    \includegraphics[width=\linewidth]{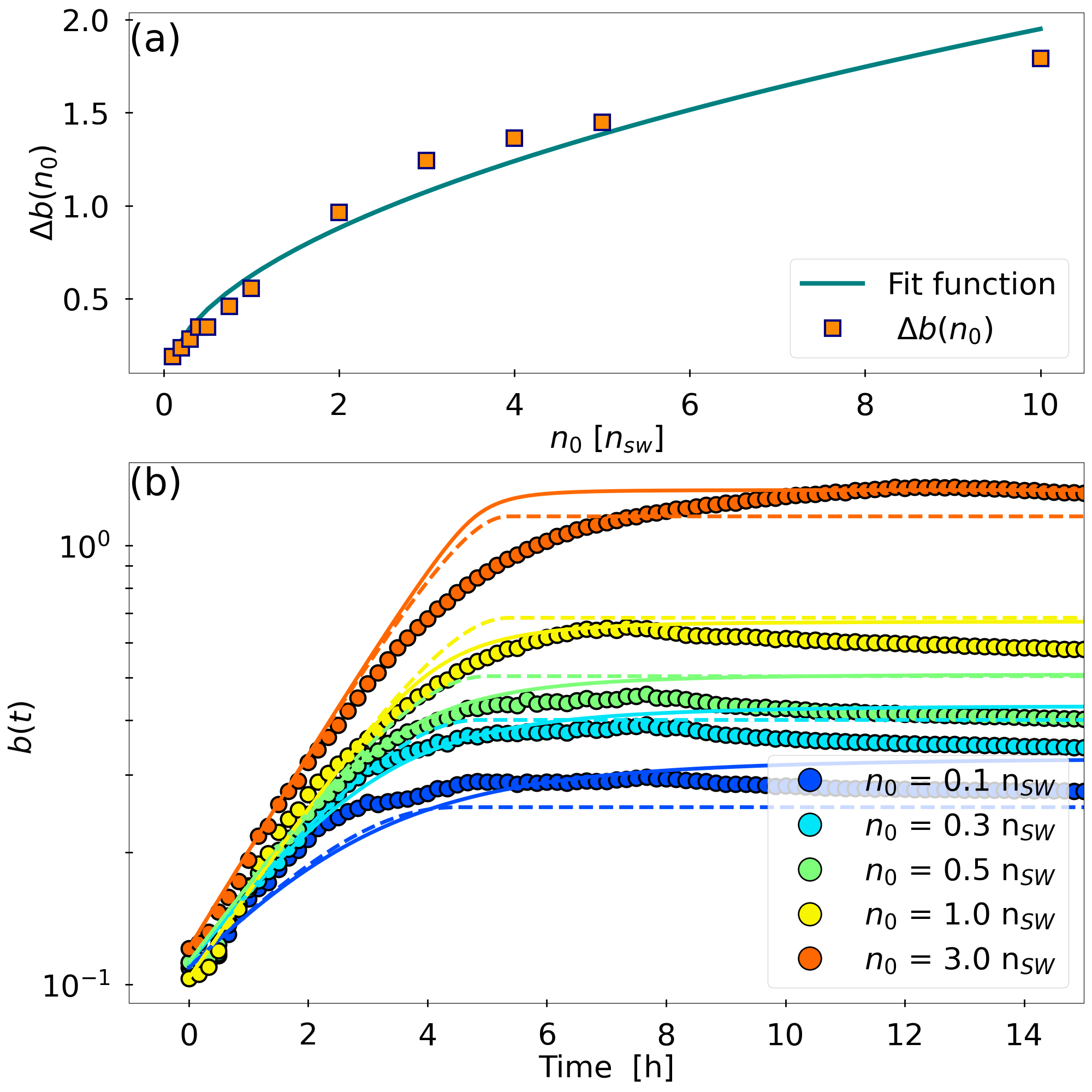}  
    \caption{{\bf Estimating nutrient consumption rate from growth curves.}
a) Dependence of biomass gain on nutrient concentration. Orange squares: experimental measurements. Dot-dashed line: fit to Eq.~\ref{eq:gain_function}.
b) Biomass growth dynamics for different initial nutrient concentrations. Disks: experimental data ($n_0=0.1, 0.3, 0.5, 1, 3$ respectively for blue, cyan, green, yellow, red). Solid lines: fits to Eqs.~\ref{eq:homogeneus_system}. Dashed lines: model dynamics resulting from the fit in panel a). Both cases assume $\gamma(n)$ as per Fig.~\ref{fig_06}b.
    }
    \label{fig_07}
\end{figure}

A direct estimate of $\alpha(n)$ requires monitoring nutrient consumption, a difficult feat in complex media. Within the approximation that treats the nutrients' pool as a single entity, the simplest approach would be to assume a constant biomass yield per nutrient consumed. This results in a biomass $b(t, n_0)$ which necessarily saturates at a level $b_{\text{sat}}(n_0)$ proportional to the initial nutrient concentration $n_0$ \cite{meyer15}. As we confirm also here (Fig.~\ref{fig_07}a), this is usually not the case. The biomass yield generally decreases with increasing nutrient concentration, a phenomenon often referred to as ``waste from haste'' \cite{pfeiffer01,gudelj10,bonachela11,meyer15,nev21}. Here, instead of modelling directly biomass yield, we generalise the functional dependence of the nutrient consumption rate $\alpha$ on $n$ to a Hill function:
\begin{equation}
\alpha(n) = \frac{n^a}{\nu_{\alpha}^a+n^a}\alpha_{\text{max}}.
\label{eq:alpha}
\end{equation}
The unknown parameters are then fixed by fitting the experimental growth curves to those obtained numerically from Eqs.~\ref{eq:homogeneus_system},\ref{eq:growth_rate},\ref{eq:alpha}. This procedure yields $\alpha_{\text{max}} = (0.0289\pm0.0003)\,\nSW \text{OD}^{-1}\text{min}^{-1}$, $\nu_{\alpha}=(0.29\pm0.01)\,\nSW$ and $a=0.92\pm 0.01$. Figure~\ref{fig_07}c shows that the best fits obtained from this simple model (solid lines) recapitulate well the experimental data.
Although in the following we will use these parameters for the heuristic description of $\alpha(n)$, similar values can be obtained in a simpler way. Equation~\ref{eq:homogeneus_system} implies a biomass gain $\Delta b(n_0)= b_{\text{sat}}(n_0)-b(0, n_0)$ given by
\begin{equation} 
    \Delta b(n_0)  =  \int_{0}^{n_0} \frac{\gamma(n(t))}{ \alpha(n(t))} dn
\label{eq:gain_int}
\end{equation} 
which, for $\gamma$ ad $\alpha$ given by Eqs.~\ref{eq:growth_rate},\ref{eq:alpha}, can be explicitly integrated to read   
\begin{align}
\begin{split}
&\Delta b(n_0) =  \frac{\gamma_{\text{max}}}{\alpha_{\text{max}}} 
\bigg[ {_2}F_1\left(1,1+\frac{1}{b},2+\frac{1}{b},-\frac{n_{0}^b}{\nu_{\gamma}^b}\right)\frac{n_{0}^{b+1}}{(b+1)\nu_{\gamma}^b} +  \\
& +{_2}F_1\left(1,\frac{1+b-a}{b},\frac{1+2b-a}{b},-\frac{n_{0}^b}{\nu_{\gamma}^b}\right)\frac{n_{0}^{1+b-a}\nu_{\alpha}^a}{(1+b-a)\nu_{\gamma}^b} \bigg].
\end{split}
\label{eq:gain_function}
\end{align}
Here  $_2F_1(\cdot)$ is the hypergeometric function. Direct fits of this analytical expression to the experimental biomass gains return values which are broadly in line with those obtained from the full growth dynamics, but without requiring the numerical integration of Eqs.~\ref{eq:homogeneus_system}  (see Table~\ref{table:alphamax} and Fig.~\ref{fig_07}c dashed lines).

\begin{table}[h]
\small
\setlength{\tabcolsep}{3pt}
\centering
\begin{tabular}{lccc}
\hline
 & $\alpha_{\max}$ 
 & $\nu_{\alpha}$ 
 & $a$ \\
 & $[n_{\mathrm{SW}}\,OD^{-1}\mathrm{min}^{-1}]$
 & $[n_{\mathrm{SW}}]$ &  \\
\hline
\makecell[l]{Theory} 
       & $0.026\, \pm 0.0012$
        & $0.9\, \pm 0.1 $
        & 1 \\
\hline
\makecell[l]{Numerical} 
       & $ 0.0289 \pm 0.0003$
        & $0.29 \pm 0.01$
        & $0.92 \pm 0.01$\\
\hline
\end{tabular}
\caption{Table reporting the final set of fitting parameters for the theoretical approach, applied to two different initial interval ranges, and for the numerical procedure.}
\label{table:alphamax}
\end{table}

\subsection{The bacteria-nutrients model replicates experimental DLs}
\begin{figure}
    \includegraphics[width=1.\linewidth]{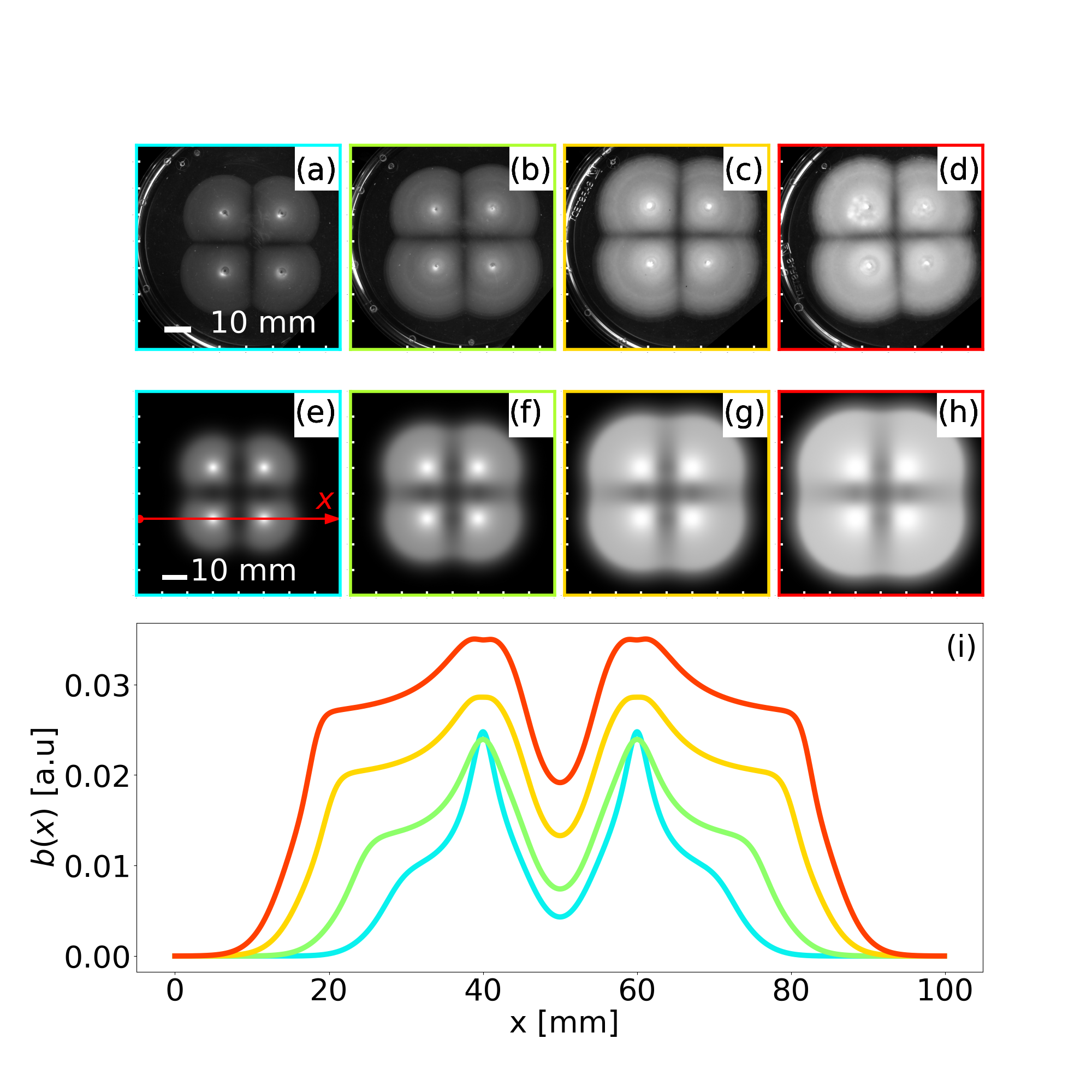}
    \caption{{\bf The minimal model recapitulates experimental observations.}
a-d) Experimental sibling inhibition patterns $24\,$h post inoculation for $n_0=0.5, 1, 2, 5$.
e-h) Simulated concentration profiles at a simulation time corresponding to $24\,$h. The set of nutrient levels is the same as in panels a-d).
i) Density profiles along the transects indicated in panel e) for $n_0=0.5, 1, 2, 5$ (cyan, green, yellow and orange respectively).
}
\label{fig_08}
\end{figure}

Armed with the experimentally validated functions $D(n)$, $\gamma(n)$  and $\alpha(n)$, and assuming a nutrient diffusivity of  $D_n = 500\,\mu$m$^2 /$s \cite{cremer19}, we can now simulate the evolution of the model in Eqs.~\ref{eq:full_system} for a given distribution of inoculation sites and initial background nutrient concentration $n_0$. 
Figure~\ref{fig_08} shows a comparison between the experimental (panels a-d) and simulated (panels e-h) cell concentration profiles for $n_0=0.5, 1, 2, 3$ at $24\,$h post-inoculation, together with transects (Fig.~\ref{fig_08}i). 
The simulations show clearly that nutrient depletion is sufficient to generate the demarcation lines observed in our experiments.

\begin{figure}[t]
    \centering
    \includegraphics[width=\linewidth]{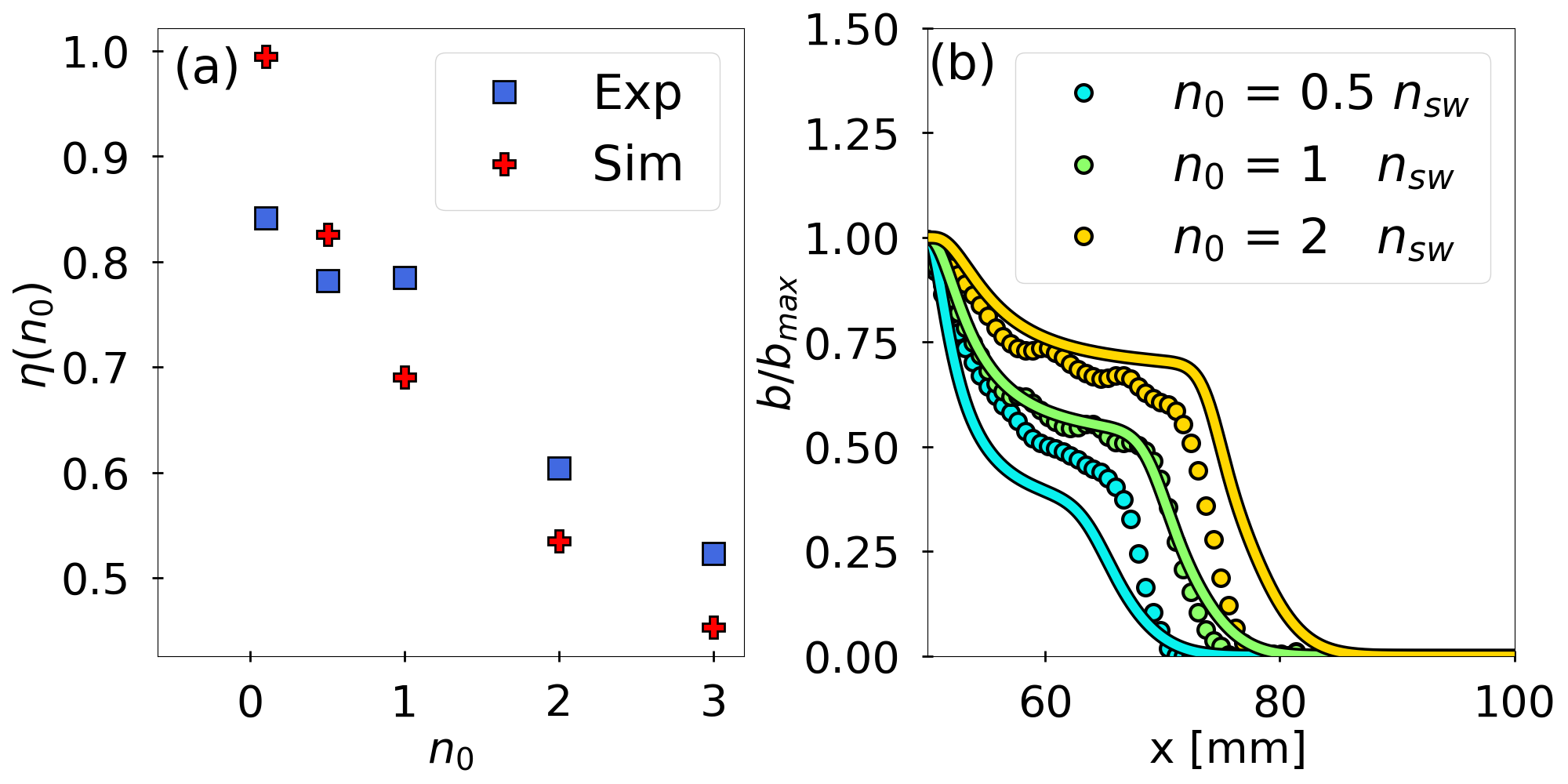}
    \caption{{\bf Comparison between experiments and numerical results.}
a) Dependence of the depth ratio $\eta(n_0)$ on initial nutrient concentration. Blue squares: experimental measurements. Red crosses: numerical predictions.
b) Normalised density profiles for single colony expansion at 20 h post inoculation, for $n_0=0.5, 1, 2$ (cyan, green, yellow respectively). Disks: experimental measurements. Solid lines: model predictions. Comparison between experimental (dots) and simulated (solid lines) normalised density profiles.
    }
    \label{fig_09}
\end{figure}

A simple but robust quantification of the effect of $n_0$ on the depth of DLs is given by the ratio
\begin{equation}
\eta(n_0) =\frac{b_{\mathrm{{max}}}(n_0)-b_{{\mathrm{{min}}}}(n_0)}{b_{\mathrm{{max}}}(n_0)},
\label{eq:eta}
\end{equation}
where $b_{\mathrm{{max}}}$ and $b_{\mathrm{{min}}}(n_0)$ are the bacterial densities at the inoculation point and in the inhibited region respectively. Figure~\ref{fig_09}a shows that the experimental values for $\eta(n_0)$  are in good agreement with those obtained from the model ($24\,$h post-inoculation). 
The model captures well also the experimental expansion profiles of individual colonies (Figure~\ref{fig_09}b), especially at intermediate values of the background nutrient concentration. The difference between the two increases with $n_0$, potentially indicating that supplementary $n_0$-dependent processes need to be taken into account to reproduce quantitatively the observed expansion dynamic. 

\section{Conclusions}
In this study, we investigated sibling inhibition in spreading bacterial colonies. Our experiments with {\it Pseudomonas aeruginosa} in soft agar hydrogels demonstrate that isogenic colonies of swimming cells form sharp demarcation lines when growing and expanding towards each other. Demarcation lines in our system are not due to lethal inhibition, gel matrix compression or quorum sensing, as previously proposed for other species. Instead, by experimentally quantifying the effect of nutrient concentration on replication rate, active diffusion and nutrient uptake, we showed that a minimal growth-motility model  is sufficient to reproduce the colonies' expansion dynamic. This includes the emergence of demarcation lines and their dependence on the initial background nutrient concentration. 
In {\it P. aeruginosa}, therefore, sibling inhibition appears to be a direct consequence of the effect that localised nutrient depletion has on replication and motility of single cells.

Understanding how microbial communities organise and interact in physically constrained environments is critical for a wide range of applications, from combating antibiotic-resistant pathogens in human tissues to engineering useful microbiomes in soil and bioreactors.  Our findings provide a simple example of how the spatial ecology of microorganisms can be shaped by the interplay between metabolic and physical factors.
The minimal model, together with the experimental validation of its individual components, provides a robust modelling approach  that can be applied to other bacterial species and porous media. 
We hope that future research will extend this framework to investigate other forms of bacterial interactions like competition and cooperation, and to explore the effects of different physico-chemical environments on the spatial distribution of bacteria.

\section{Acknowledgments}
We thank Davide Michieletto for his generous support in the gel compression experiments, Giordano Rampioni for providing us with the quorum sensing mutants of {\it P. aeruginosa} PAO1, Roberto Di Leonardo and Giacomo Frangipane for their support in developing the imaging set-up.\\FI, FB, DB and BC  would like to acknowledge the Grant of Excellence Departments, MIUR-Italy (ARTICOLO 1, COMMI 314 - 337 LEGGE 232/2016), the Rome Technopole Project (CUP:F83B22000040006). MP acknowledges the fact that IMEDEA is an accredited `Mar\'ia de Maeztu Excellence Unit' (grant CEX2021-001198, funded by MCIN/AEI/10.13039/ 501100011033).

\newpage

\section{Supplementary Materials\label{sec:SuppMat}}
\section{Calibration of the Newly Developed Imaging Setup}
We report the calibration curve for our custom-built imaging setup. The $y$-axis represents the collected signal intensity, expressed in grayscale values (arbitrary units), while the $x$-axis reports the bacterial density independently measured with a spectrophotometer as optical density at 600~nm (OD$_{600}$). The calibration data show a clear linear relationship between signal intensity and bacterial density within the interval $y \in [0,100]$. This linear regime defines the reliable operating range of the imaging system, in which grayscale intensity can be directly used as a quantitative proxy for cell density. Outside this interval, deviations from linearity are observed, likely due to detector saturation at high signal levels and reduced sensitivity at low intensities. For this reason, all measurements and analyses presented in this work are restricted to the range $y \in [0,100]$, ensuring accuracy and reproducibility.
\begin{figure}[hb]
    \centering
\includegraphics[width=0.5\linewidth]{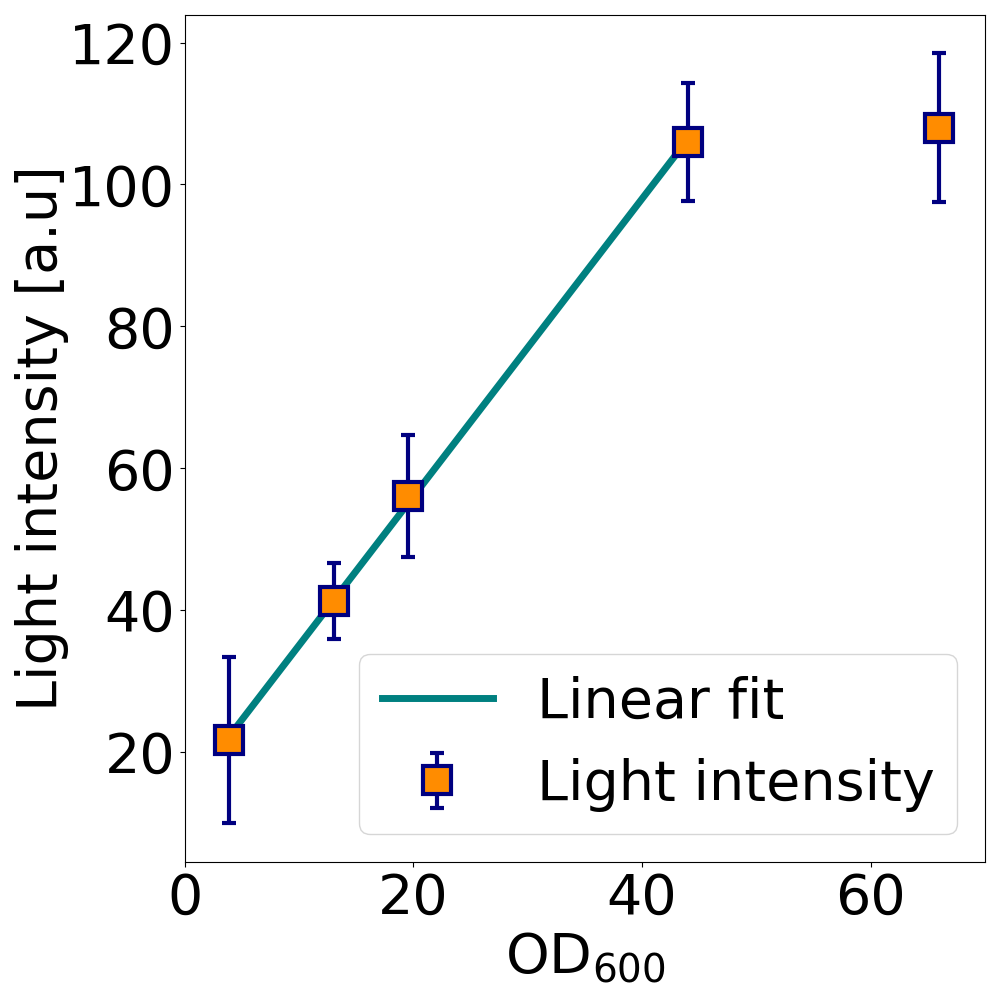}
 \caption{
Calibration curve for the custom-built imaging setup. Grayscale signal (arb. units) is plotted against bacterial density (OD$_{600}$). A linear correlation is observed for $y \in [0,100]$, which defines the range used in this work.} 
\label{fig:S1}
\end{figure}

\section{Measuring growth and consumption rates }

The system in the 96-well plate evolves according to 
\begin{equation}
\begin{cases} 
\partial_t b(t) =  \gamma(n(t))\,b(t),\\ 
\partial_t n(t) = - \alpha(n(t))\,b(t).
\end{cases}
\end{equation}

We begin the analysis by focusing on the initial portion of the $b(t)$ curves, arguing that for sufficiently short times $n(t)$ can be approximated as constant. The validity of this approximation depends on the relative change of $n(t)$ with respect to its initial value over the considered time interval.

Defining $n(t)=n_0(1-\epsilon(t))$ and expanding $\gamma(n)$ and $\alpha(n)$ around $\epsilon=0$, we obtain
\begin{equation}
\begin{cases} 
\partial_t b(t) =  \left(\gamma(n_0)-n_0\frac{\partial\gamma}{\partial n}\big|_{n_0}\epsilon +\mathcal{O}(\epsilon^2)\right)b(t),\\ 
\partial_t \epsilon(t) =   \left( \frac{\alpha(n_0)}{n_0} -\frac{\partial\alpha}{\partial n}\big|_{n_0}\epsilon+\mathcal{O}(\epsilon^2)\right) b(t).
\end{cases}
\label{eq:rates_taylor}
\end{equation}

\begin{figure}[h!]
    \centering
    \includegraphics[width=0.5\linewidth]{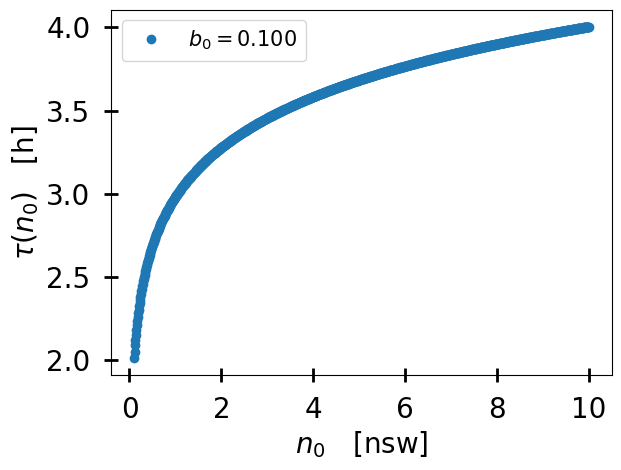}
    \caption{We here report the value $\tau(n_0)$ vs $n_0$ obtained for the initial concentration $b_0=0.1$}
    \label{SI_tau}
\end{figure}

To retain only the initial growth rate in the equation for $b(t)$, we require
\[
\left|n_0\frac{\partial\gamma}{\partial n}\big|_{n_0}\epsilon\right|\ll \gamma(n_0),
\]
which implies $\epsilon \ll 1$. In this regime, the equation for $\epsilon$ can also be truncated at first order. Assuming $b(t)=b_0 e^{\gamma(n_0)t}$, we obtain
\begin{equation}
\epsilon(t) = \frac{\alpha(n_0)b_0}{\gamma(n_0)n_0}\left(e^{\gamma(n_0)t} -1\right),
\label{eq:epsilon}
\end{equation}
which leads to the condition
\begin{equation}
t \ll \tau(n_0) = \frac{1}{\gamma(n_0)}\log\left(\frac{\gamma(n_0)n_0}{\alpha(n_0)b_0} +1\right).
\label{eq:tau}
\end{equation}
We report in figure \ref{SI_tau} the functional form of $\tau(n_0)$ as a function of the initial concentration $n_0$ obtained for a selected initial bacterial concentration $b_0$.

Using the fitted forms of $\gamma(n)$ and $\alpha(n)$ from the main text, with parameters $\nu_{\gamma}=0.12$, $\gamma_{\max}=0.6 \,\,h^{-1}$, $b=0.5$,  $\nu_{\alpha}=5.5\times10^{10}$, $\alpha_{\max}=1.2\times 10^{7}$, and $a=0.64$, and taking $b_0=0.1$, we estimate $\tau(n_0)\sim 2\,\mathrm{h}$ for $n_0=0.1$ and $\tau(n_0)\sim 4\,\mathrm{h}$ for $n_0=10$. Although the precise meaning of ``$\ll$'' is context-dependent, this suggests that fitting over time windows of several hours may already probe the regime where deviations from pure exponential growth become relevant. This is consistent with the downward curvature observed in semi-logarithmic plots, which reflects the increase of $\epsilon(t)$ over time.

A first correction beyond the constant-$n$ approximation can be obtained by inserting Eq.~\eqref{eq:epsilon} into the equation for $b(t)$:
\begin{equation}
b(t) \simeq b_0 \exp\left[\int_0^t \left( \gamma(n_0)-n_0\frac{\partial\gamma}{\partial n}\big|_{n_0}\epsilon(t') \right)dt'\right],
\end{equation}
which yields
\begin{equation}
b(t) \simeq b_0 e^{\gamma(n_0)t}\exp\left[-\frac{\partial \gamma}{\partial n}\big|_{n_0}\frac{\alpha(n_0)b_0}{\gamma^2(n_0)}\left(e^{\gamma(n_0)t}-(1+\gamma(n_0)t)\right)\right].
\end{equation}

Expanding for short times, we obtain the approximate form
\begin{equation}
\label{eq:b_approx}
b(t) \simeq b_0 e^{\gamma(n_0)t}\exp\left[-\frac{1}{2}\alpha(n_0)b_0\frac{\partial \gamma}{\partial n}\big|_{n_0} t^2\right],
\end{equation}
which naturally explains the observed downward curvature in semi-logarithmic plots.

This suggests an improved fitting strategy: performing a quadratic fit of $\log b(t)$ allows one to extract both $\gamma(n_0)$ (linear term) and a combination involving $\alpha(n_0)$ and $\partial_n \gamma(n_0)$ (quadratic term). Once $\gamma(n)$ is determined, its derivative can be used to estimate $\alpha(n_0)$, providing a self-consistent way to constrain the consumption rate.
We can thus perform a polynomial fit of 
\begin{equation}
\label{eq:log_b}
\ln b(t) \simeq \ln b_0 + \gamma(n_0)\, t -\frac{1}{2}\alpha(n_0)b_0\frac{\partial \gamma}{\partial n}\Big|_{n_0} t^2,
\end{equation}
so that $\ln b(t)$ is approximated by a quadratic polynomial, $\ln b(t) = A + B\, t + C\, t^2$, with coefficients
\begin{equation}    
\begin{cases}
\label{eq:poly_fit_coefficient}
A = \ln b_0, \\
B = \gamma(n_0), \\
C = -\dfrac{1}{2}\alpha(n_0)b_0\frac{\partial \gamma}{\partial n}\Big|_{n_0}.
\end{cases}
\end{equation}

Once the coefficients $A$, $B$, and $C$ are determined, we obtain a direct estimate of the replication rate $\gamma(n_0)$ from the linear term. 

\begin{figure}[hb]
    \centering
    \includegraphics[width=.8\linewidth]{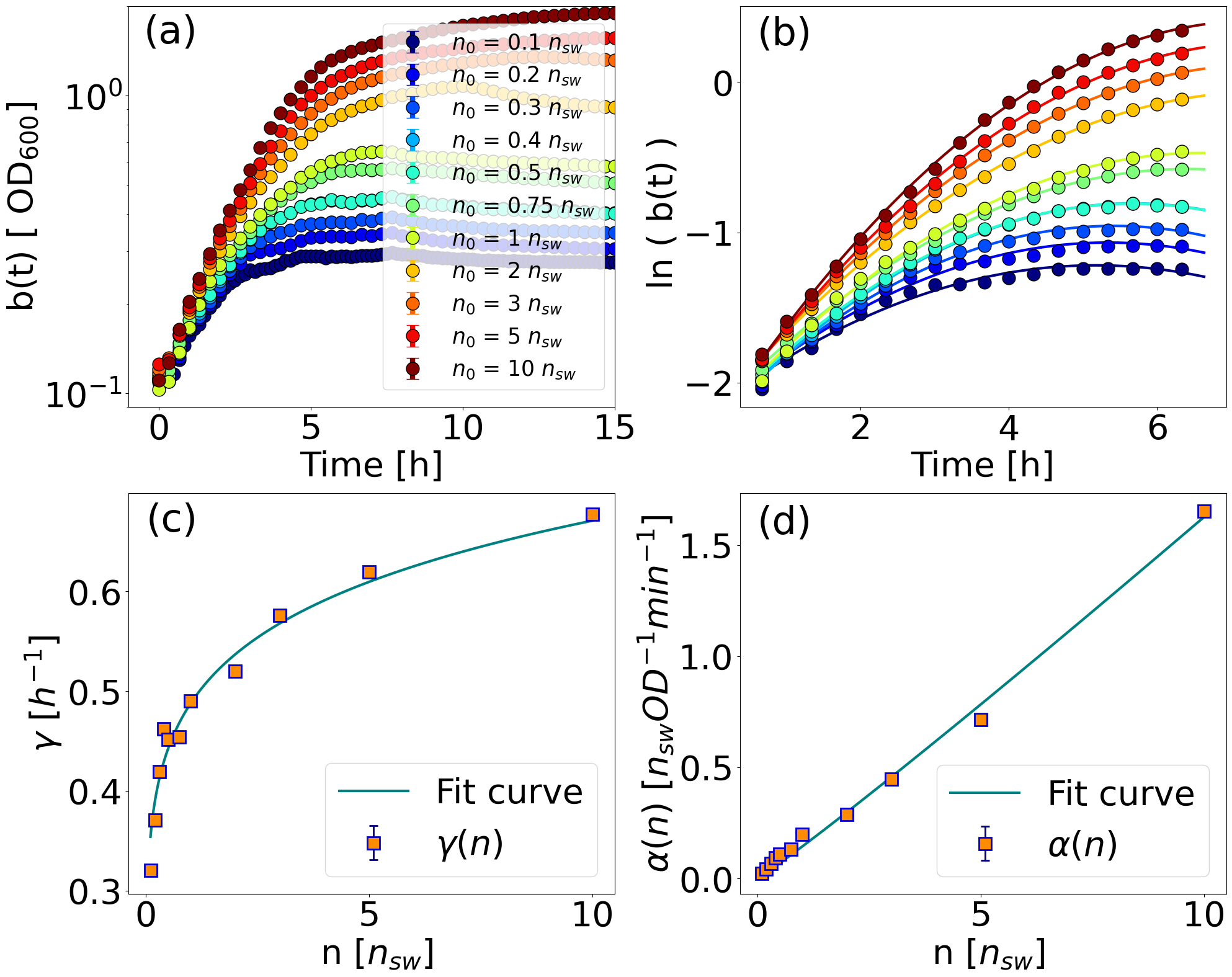}
    
    \caption{\textbf{Replication rates estimated  with polynomial fitting of $\mathbf{ln \, b(t)}$. }\\
    \textbf{Panel (a):} Growth of PAO1 in liquid medium. Bacterial density, measured as optical density OD${600}$ (light absorbance at 600 nm), is shown as a function of time over a 15-hour interval. Each color corresponds to a different nutrient concentration in the growth medium, spanning from $10^{-1},n_\mathrm{SW}$ to $10^{1},n_\mathrm{SW}$. The value $1,n_\mathrm{SW}$ denotes the reference concentration used for the standard swimming assay described in the Methods section.\\
    \textbf{Panel (b):} Log-transformed data points within the fitting interval and corresponding second-degree polynomial fits, as described in Eq.~\ref{eq:log_b}.\\
    \textbf{Panel (c):} Replication rate $\gamma(n_0)$ estimated from the polynomial fit as the first-degree coefficient $B$ (Eq.~\ref{eq:poly_fit_coefficient}). The teal curve represents a Hill function fit:
     $\gamma(n) = \gamma_{\text{max}} \frac{n^b}{\nu_{\gamma}^b+n^b}$.
    \textbf{Panel (d):} Consumption rate $\alpha(n_0)$ estimated - although no clear saturation is observed within the explored range of $n_0$ - from the polynomial fit as the second-degree coefficient $C$ (Eq.~\ref{eq:poly_fit_coefficient}). The teal curve represents a Hill function fit:
    $\alpha(n) = \alpha_{\text{max}} \frac{n^a}{\nu_{\alpha}^a+n^a}$
    }
      
    \label{fig:S2_gamma_alpha}
\end{figure}

Result of this fitting procedure are reported in Fig.~\ref{fig:S2_gamma_alpha}, panel~(c). 

\begin{figure}[hb]
    \centering
    \includegraphics[width=.8\linewidth]{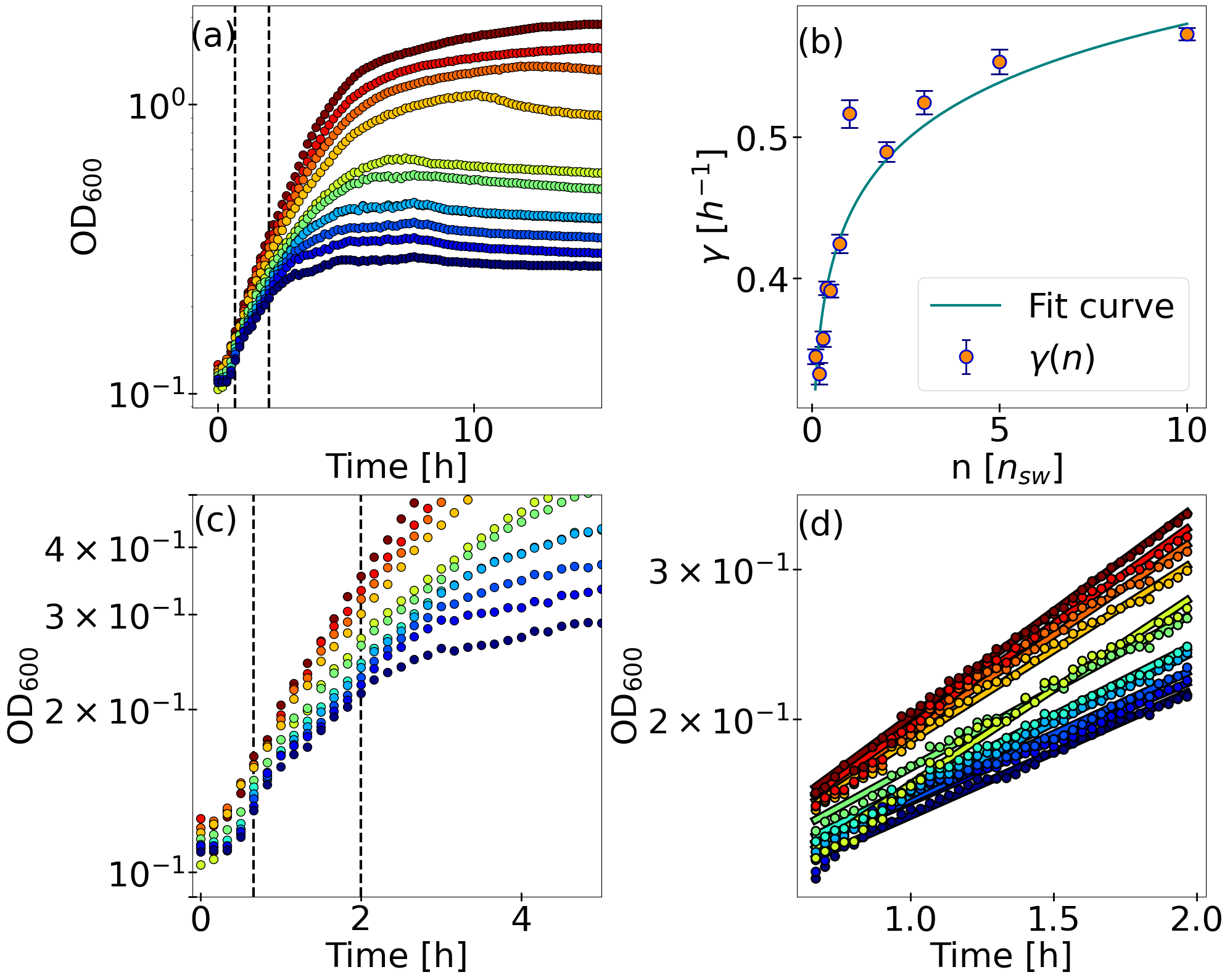}
    \caption{\textbf{Replication rates estimated with an exponential fit.}\\
    \textbf{Panel (a):} Growth of PAO1 in liquid medium. Bacterial density, measured as optical density OD${600}$ (light absorbance at 600 nm), is shown on a semi-log scale as a function of time over a 15-hour interval. Each color corresponds to a different nutrient concentration in the growth medium, spanning from $10^{-1} n{\mathrm{SW}}$ to $10^{1} n_{\mathrm{SW}}$. The value $1,n_{\mathrm{SW}}$ denotes the reference concentration used for the standard swimming assay described in the Methods section. Dashed vertical lines indicate the selected fitting interval.\\
    \textbf{Panel (b):} Replication rates $\gamma(n_0)$ estimated from an exponential fit of the form $b(t) = b_0 e^{\gamma(n_0)t}$. The teal curve represents a Hill function fit:
        $\gamma(n) = \gamma_{\text{max}} \frac{n^b}{\nu_{\gamma}^b + n^b}$.
       \textbf{Panel (c):} A close-up of the first 5 hours of growth highlights the presence of a lag phase (approximately the first 30 minutes) and supports the choice of the fitting interval $t \in [1,2]$.\\ 
    \textbf{Panel (d):} Data points within the fitting interval are shown on a semi-log scale together with their exponential fits. The linear trend confirms the expected behavior of true exponential growth.
    }
    \label{fig:S2_simple_gamma}
\end{figure}

Notably, this estimate is in very good agreement with that obtained from a simple exponential fit, shown in Fig.~\ref{fig:S2_simple_gamma}, panel~(b), despite the absence of any constraint enforcing such consistency.

\begin{figure}[hb]
\centering  
\includegraphics[width=\linewidth]{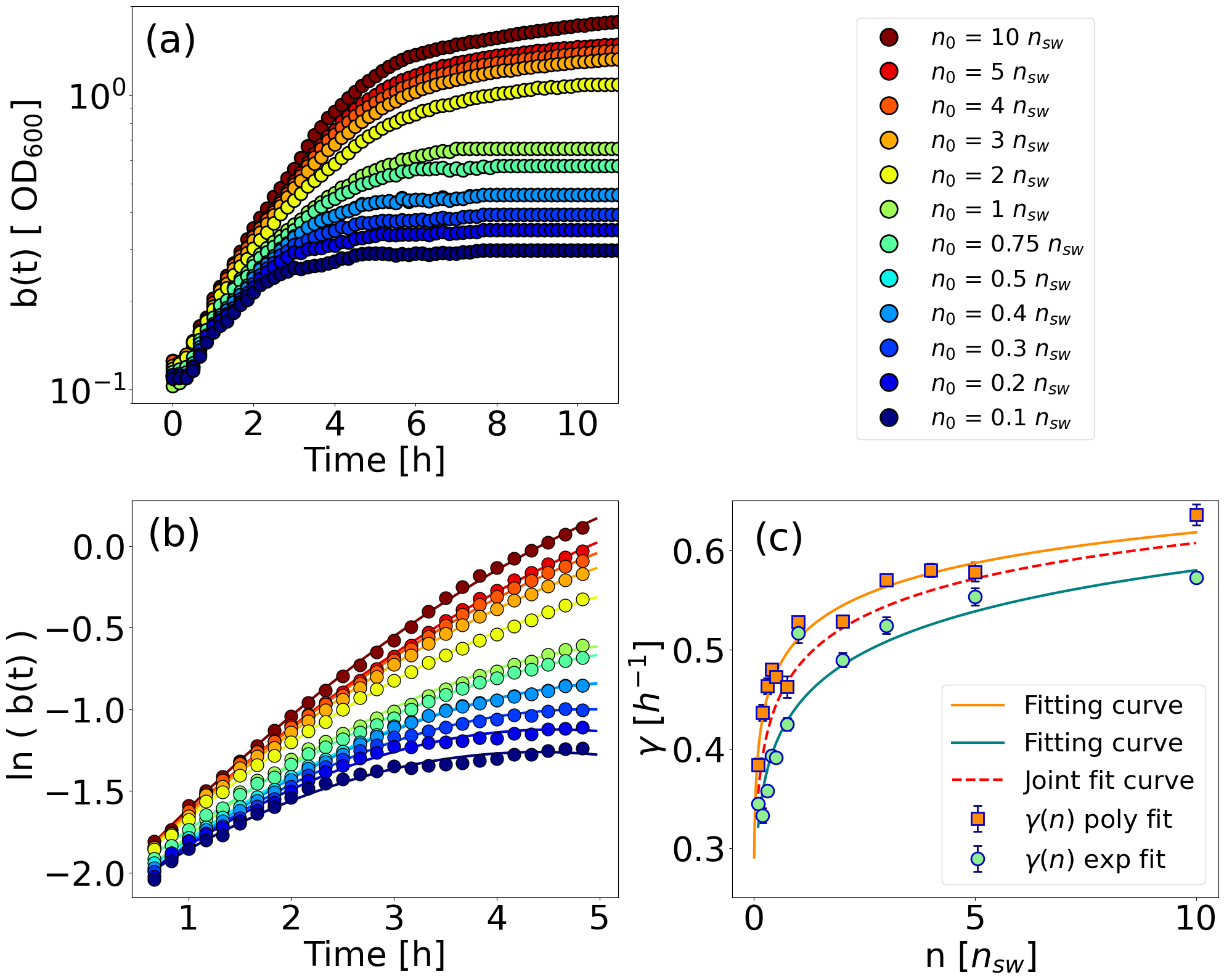}
\caption{\textbf{Growth dynamics of PAO1 for different nutrient concentrations.}
\textbf{Panel (a):} Growth curves of \textit{P. aeruginosa} for various initial nutrient concentrations, showing that higher nutrient concentration results in faster growth and higher saturation density.
\textbf{Panel (b)} Natural logarithm of the growth phase of the  curves reported in panel (a) together with their fitting results as in eq Eq. \ref{eq:log_b} .
\textbf{Panel (c):} Replication rate $\gamma(n)$ as a function of the initial nutrient concentration. Blue dots data points are obtained by fitting the exponential phase of the  curve in panel (a) corresponding to the same $n_0$ value. Orange square data points are obtained with the polynomial fitting procedure described in the main text.
The exponential fitting yields slightly lower replication rates for each nutrient level but overall the two fitting procedure are comparable.
Blue and orange solid lines are the results of  fitting, blue and orange data points respectively) to $\gamma(n)=\gamma_{max} \,\, n^b/(\nu^b + n^b)$.
Red dashed lines fits all data points together.
}
\label{fig_S5}
\end{figure}

We report the comparison between the results obtained with the two fitting procedures in panel c of Fig.~\ref{fig_S5}. 

Once an estimate of the replication rate is obtained, it can be fitted with a Hill function, allowing us to compute its derivative $\partial \gamma / \partial n$ over the full range $n_0 \in [0.1, 10]$. Using this, the consumption rate can be estimated as
\[
\alpha(n_0)= \frac{2C}{b_0 \,\frac{\partial \gamma}{\partial n}\big|_{n_0}},
\]
with the results reported in Fig.~\ref{fig:S2_gamma_alpha}, panel~(d).

For consistency, $\alpha(n)$ is also fitted with a Hill function. However, since no clear saturation is observed within the explored range $n_0 \in [0.1, 10]$, the parameters $\alpha_{\max}$ and $\nu_{\alpha}$ should not be overinterpreted. Despite this limitation, the fitting procedure provides valuable information, as it enables us to simulate growth dynamics using the inferred form of $\alpha(n)$.

\section{Diffusion simulations}
Once defined the model and calibrated    $D( n)$, $\gamma( n)$  and $\alpha( n)$ on the experimental data (for the latter we will make use of the values obtained through the fully numerical procedure), we can perform simulations of bacterial growth and expansion.  
Following the procedure described in the methods section, we perform an extensive set of simulations to reproduce the expansion of multiple colonies growing within the same media for different nutrient levels $n_0$. 
To this aim, we set our simulations to start on a grid of 1000x1000 pixels, where each pixel corresponds to 100 $\mu$m.
We aim at reproducing the experiments where 4 colonies are inoculated at a distance  $d=20$ mm from one each other, and their growth is analysed in time, for different initial nutrient levels $n_0$.
To this aim we initialise our dynamics selecting 4 points at a relative distance $d= 200$ pixels on a square pattern whose centre corresponds to the centre of  the grid.
These 4 points will correspond to the inoculating spot of each of the 4 sibling bacterial colonies. 
As a starting condition, we assume that each bacterial colony starts as a narrow Gaussian distribution ($\sigma=500\mu$m) centred around the inoculation point. 
Initially every cell of the grid has the same $n_0$ value.
Throughout the simulation we assume a random (thermal) diffusion for the nutrient with diffusion constant   $D_n = 500\,\mu$m$^2 /$s, whose order of magnitude is consistent with estimates found in literature~\cite{cremer19}.
Simulations are then performed by using all numerical values - together with their original units of measurements - for diffusion, consumption and replication rates as obtained  calibrating the model on  experimental data.
Thus all comparison between the numerical results and the experimental data has to be considered in experimental units (mm for lengths, s for time, if not otherwise stated).

\section{Comparison between It\^o and anti-It\^o  conventions}

The extension of the FKPP reaction term  to include the space- and time-dependent expression $\gamma\left(n(x,t)\right ) b(x,t)$ is conceptually straightforward. However, its practical implementation necessitates a rigorous investigation into the functional forms of $\gamma\left(n(x,t)\right)  $ and $\alpha\left(n(x,t)\right ) $ with respect to the nutrient density $n(x,t)$.\\
In contrast, extending the diffusion term to incorporate a space-dependent diffusivity, $D(x)$, is a non-trivial matter.
A robust derivation requires starting from a microscopic Langevin equation, which, depending on the chosen convention (or interpretation), yields distinct equivalent Fokker-Planck equations. 
The most common interpretations are the It\^o, Stratonovi\v{c}, and the isothermal (or anti-It\^o) conventions, leading to three structurally distinct forms for the diffusion term such as  $\nabla^2\left(Db\right)$ (It\^o),  $\nabla\cdot\left(D\nabla b\right)$ (Anti-It\^o or Isothermal) or $\nabla\cdot\left(\sqrt{D}\;\nabla\sqrt{D}b\right)$ (Stratonovi\v{c})\cite{gardiner2009}. 
Despite the apparent structural differences, the physical results derived from these equations remain consistent, as established equivalence relationships allow for precise mapping between the coefficients of each formulation \cite{gardiner2009}.\\
The choice among the derived Fokker-Planck forms is critical, as simply prescribing a space-dependent diffusivity $D(x)$ and using one of these structural forms will result in drastically different evolutions of the probability density $b(x,t)$. For instance, in the absence of a reaction term, the It\^o interpretation, $\nabla^2\left(Db\right)$, generates a stationary distribution $b_\mathrm{{stat}}(x) \propto 1/D(x)$, whereas the anti-It\^o interpretation, $\nabla\cdot\left(D\nabla b\right)$, yields a uniform stationary distribution $b_\mathrm{{stat}}(x) = \text{const}$.\\
Consequently, without precise knowledge of the underlying microscopic dynamics (specifically, all coefficients of the associated Langevin equation) it is impossible to deduce the correct form of the diffusion term. The mere enforcement of a particular space-dependent function $D(x)$ is therefore insufficient to fully define the macroscopic dynamics.\\
 Tupper and Yang\cite{tupper12} showed that distinct microscopic dynamics can generate an identical macroscopic diffusivity $D(x)$, yet yield completely different stationary probability distributions $b_\mathrm{{stat}}$. 
In the context of an active fluid bath where the self-propulsion speed is spatially dependent, $v(x)$, theoretical and experimental evidence suggests that the stationary density distribution, $b_\mathrm{{stat}}$, will not be uniform. Instead, it is generally expected to be inversely proportional to the propulsion speed: $b_\mathrm{{stat}}(x) \propto 1/v(x)$ ~\cite{arlt18,frangipane18,pellicciotta23}.\\
Since the average diffusivity of self-propelling particles is typically proportional to their self-propulsion speed, the observed stationary state $b_\mathrm{{stat}}(x) \propto 1/D(x)$ corresponds to the steady-state solution generated by the It\^o interpretation of the Fokker-Planck equation, $\nabla^2\left(Db\right)$. Therefore, in the absence of complete microscopic information, we consider the It\^o formulation to be a reasonable choice for modelling the dynamics in this study.

\bibliography{biblio}
\bibliographystyle{apsrev4-2}

\end{document}